\documentclass[12pt,a4paper,psamsfonts]{article}
\usepackage{amsmath,amssymb,amsthm}
\usepackage{eucal}

\numberwithin{equation}{section}

\usepackage{geometry}

\def\d{\delta}

\def\CC{\mathbb C}

\def\R{{\Bbb R}}

\newcommand{\be}{\begin{equation}}
\newcommand{\ee}{\end{equation}}
\newcommand{\bea}{\begin{eqnarray}}
\newcommand{\eea}{\end{eqnarray}}

\newcommand{\al}{\alpha}
\renewcommand{\d}{\delta}
\newcommand{\e}{\epsilon}
\newcommand{\G}{\Gamma}
\newcommand{\g}{\gamma}

\newcommand{\La}{\Lambda}
\newcommand{\la}{\lambda}
\newcommand{\m}{\mu}
\newcommand{\n}{\nu}
\newcommand{\Om}{\Omega}
\newcommand{\om}{\omega}
\newcommand{\s}{\sigma}

\newcommand{\hlf}{\frac{1}{2}}

\newcommand{\rr}{\rightarrow}
\newcommand{\w}{\wedge}
\newcommand{\Z}{\mathbb{Z}}

\newcommand{\SO}{\operatorname{SO}}

\newcommand{\Spin}{\operatorname{Spin}}
\newcommand{\SU}{\operatorname{SU}}

\newcommand{\lp}{\left(}
\newcommand{\rp}{\right)}
\newcommand{\ls}{\left[}
\newcommand{\rs}{\right]}

\newcommand{\hph}[1]{{\hphantom{#1}}}
\newcommand{\non}{\nonumber}
\newcommand{\ul}[1]{{\underline{#1}}}

\makeatletter
\renewcommand\section{\@startsection {section}{1}{\z@}%
                                   {-3.5ex \@plus -1ex \@minus -.2ex}
                                   {2.3ex \@plus.2ex}%
                                   {\normalfont\large\bfseries}}
\renewcommand\subsection{\@startsection{subsection}{2}{\z@}%
                                     {-3.25ex\@plus -1ex \@minus -.2ex}%
                                     {1.5ex \@plus .2ex}%
                                     {\normalfont\bfseries}}
\makeatother

\begin{document}

\begin{center}
\addtolength{\baselineskip}{.5mm}
\thispagestyle{empty}
\begin{flushright}
{\sc MIFPA-14-38}\\
\end{flushright}

\vspace{20mm}

{\Large  \bf String Corrected Spacetimes and $\SU(N)$-Structure Manifolds}
\\[15mm]
{Katrin Becker, Melanie Becker, and Daniel Robbins}
\\[5mm]
{\it George P. and Cynthia W. Mitchell Institute for }\\
{\it Fundamental Physics and Astronomy, Texas A\& M University,}\\
{\it College Station, TX 77843-4242, USA}\\[5mm]

\vspace{20mm}

\begin{center}

{\it Submitted to Nuclear Physics B Special Issue: 60 Years of Calabi Conjecture}

\end{center}

\end{center}
\abstract
Using an effective field theory approach and the language of $\SU(N)$-structures, we study higher derivative corrections to the supersymmetry constraints for compactifications of string or M-theory to Minkowski space.  Our analysis is done entirely in the target space and is thus very general, and does not rely on theory-dependent details such as the amount of worldsheet supersymmetry.  For manifolds of real dimension $n<4$ we show that internal geometry remains flat and uncorrected.  For $n=4,6$, K\"ahler manifolds with $\SU(N)$-holonomy can become corrected to $\SU(N)$-structure, while preserving supersymmetry, once corrections are included.


\vfill
\newpage

\tableofcontents

\section{Introduction}


Ever since the discovery of the relevance of Calabi-Yau manifolds in the context of string theory \cite{Candelas:1985en}
physicists have wondered if corrections (either perturbative or non-perturbative) to the internal geometry $M$ would spoil the property that K\"ahler manifolds with vanishing first Chern class $c_1(M)=0$ emerge as a solution to the theory.
The vanishing of the first Chern class is a topological condition that holds, in particular, when the K\"ahler manifold admits a Ricci-flat metric. In fact, Calabi conjectured \cite {Calabi} and Yau proved \cite{Yau:1977ms} many years ago that a compact K\"ahler manifold with $c_1(M)=0$ admits a K\"ahler metric with $\SU(N)$-holonomy in the same K\"ahler class.

Since a spinor representation of $\Spin(2N)$ always contains a singlet when the group is reduced to $\SU(N)$, one can always find a covariantly constant spinor on $\SU(N)$-holonomy manifolds
\be
\label{eq:CovConstSpinor}
\nabla_a\eta=0.
\ee
Here and in the following $a,b,\dots $ denote the coordinates on $M$.
This also implies that such geometries are Ricci-flat. To see this, note that the integrability condition for (\ref{eq:CovConstSpinor}) is
\be
\ls\nabla_a,\nabla_b\rs\eta=0.
\ee
This leads to the equation
\be
R_{abcd}\G^{cd}\eta=0,
\ee
which by contraction with $\G^b$ results in a vanishing Ricci tensor, $R_{ab}=0$.

How is this discussion related to the internal geometries and how do we take string theory corrections into account?  This is the underlying idea discussed in this paper.

The basic situation we have in mind is that we are starting with some effective space-time theory in $D$ dimensions formulated in terms of a derivative expansion.
We consider the effect that string theory corrections (either perturbative or nonperturbative) have on the vacuum solution.
More precisely, we are looking for supersymmetric solutions which have the form of a product of $(D-n)$-dimensional Minkowski space times an $n$-dimensional internal space $M$.
To find such a solution, we must ensure that the supersymmetry variations of the fermionic fields vanish for some choice of nowhere-vanishing anticommuting supersymmetry parameter $\e$.

Since we are looking for manifolds with a Minkowski space factor, it is natural to decompose the $\Spin(D-1,1)$ spinor $\e$ into $\Spin(D-n-1,1)\times\Spin(n)$ schematically\footnote{Depending on exactly what dimensions and which spinor representations are involved, we may need to include a couple of terms in this decomposition. For instance, if all the dimensions are even and $\e$ is positive chirality, then we would have $\e_+=(\xi_+\otimes\eta_+)\oplus(\xi_-\otimes\eta_-)$, where subscripts indicate chiralities.  But these subtleties will not concern us in setting up this general story.} as $\e=\xi\otimes\eta$, where $\xi$ is an anticommuting constant spinor on $\R^{D-n-1,1}$ and $\eta$ is a commuting spinor on the manifold $M$.

The vanishing of the SUSY variation of the gravitino gives, to leading order, precisely the condition (\ref{eq:CovConstSpinor}) mentioned earlier
\be
0=\d\psi_A=\nabla_A\e\quad\Longleftrightarrow\quad\nabla_a\eta=0.
\ee
Here the covariant derivative on the right is with respect to the metric on $M$ and indices with capital letters are ten dimensional.
Since we have turned off fields other than the metric, the remaining fermionic variations will automatically vanish, as will other terms in the gravitino variation\footnote{This is the simplest situation, but our techniques can be applied more broadly.  For instance, in string theory, fluxes are quantized in string units.  Thus if we turn on a fixed number of flux quanta, but can access a large volume limit (i.e.\ the volume of the internal space is large in string units), then the magnitude of the flux density scales as a positive power of $\al'$ and can be moved to subleading order in the perturbation series.  This is sometimes called the {\it{dilute flux approximation}}.  Note, however, that we should still make sure that other SUSY variations vanish, such as the dilatino variation.  We won't pursue this line of attack in the current paper.}.

If we now include higher order corrections in our effective theory, the equation we need to solve becomes
\be
\label{eq:GeneralSituation}
\nabla_a\eta=X_a[g]\eta.
\ee
Here $X_a[g]$ is some matrix\footnote{In fact, $X_a[g]$ could even in principle involve differential operators acting on $\eta$.  For instance we could imagine some higher order correction in string theory leading to a term like $(\al')^mR^m\nabla_a\eta$ appearing on the right hand side of the corrected gravitino variation (where $R^m$ is some scalar constructed from $m$ Riemann tensors).  This would make the discussion a little more involved but would not qualitatively change the situation, nor would it invalidate the approach being presented.} constructed covariantly from the metric $g_{ab}$ on $M$.
  This matrix encodes the string theory corrections, whose form depends on the dimension of $M$ and the precise theory being considered, as we see later on. In summary, we want to find a metric $g_{ab}$ on $M$, and a nowhere-vanishing spinor $\eta$, such that (\ref{eq:GeneralSituation}) is satisfied.  From our previous discussions we see that the corrected (physical) metric may no longer necessarily have a reduced holonomy group, nor is it necessarily Ricci-flat.

Here we only need $\eta$ to be nowhere-vanishing, but of course, we can always use this to construct a unit normalized spinor $\eta'=\eta/\sqrt{\eta^\dagger\eta}$ satisfying $\eta^{\prime\,\dagger}\eta'=1$.  Then $\eta'$ obeys an equation of similar form to (\ref{eq:GeneralSituation}), but with $X_a$ shifted by a term proportional to $(\eta^{\prime\,\dagger}X_a\eta')$ times the identity matrix.  In other words, $\eta'$ obeys
\be
\label{eq:NormalizedSituation}
\nabla_a\eta'=X'_a[g,\eta']\eta'.
\ee
Henceforth we will assume that we have normalized our spinor in this way, and we will drop the primes.

Our goal is to solve (\ref{eq:NormalizedSituation}), and learn about the properties of the internal geometry of $M$ along the way.  Note that this equation is very nonlinear in the metric, so solving it can be quite non-trivial.
In fact, no explicit analytic expressions for Calabi-Yau metrics are known yet (see the discussion of \cite{Douglas:2015aga} in this volume).
One reasonable approach is to proceed order by order in an expansion.  At each step the lower order solution will act as source terms to solve for the next order.



A useful alternative to solving directly for $g_{ab}$ and $\eta$ involves the construction of bilinears.  In each dimension we can construct various differential forms on $M$ by sandwiching antisymmetrized products of gamma matrices between $\eta^T$ and $\eta$ or their complex conjugates.  In fact, we can typically find an invertible mapping between $\{g_{ab},\eta\}$ and some subset of bilinears, possibly with constraints.

That is, given $g$ and $\eta$ we can of course construct our bilinears, but conversely given some subset of the bilinears, possibly obeying certain constraints, we can construct a metric and a normalized spinor (if the bilinears were known explicitly).  We will see this in more detail in each dimension below.  Using this map, we can convert the equation (\ref{eq:NormalizedSituation}) into an equivalent set of equations on these bilinears which can often be much easier to work with.  Again the details are very dimension dependent and are explored below.

Note that the existence of a normalized spinor on the manifold $M$ reduces the structure group from $\operatorname{GL}(n)$ to some smaller group $G$, i.e.\ we can construct an atlas of patches $\mathcal{U}_i$ on $M$ where the transition functions mapping between $T\mathcal{U}_i$ and $T\mathcal{U}_j$ are elements of $G\subset\operatorname{GL}(n)$.

Since $\eta$ and $g_{ab}$ are equivalent to some collection of forms - the bilinears discussed above - then this set of forms also accomplishes the same reduction of the structure group down to $G$. A specification of such a set of forms is called a $G$-structure.  Our task will be to convert (\ref{eq:NormalizedSituation}) into equations for a $G$-structure.
More concretely, we show that in M-theory/string theory the internal manifold has $SU(N)$-structure once higher order corrections to the space-time effective action are taken into account. We explicitly calculate the $SU(N)$-structure for manifolds of real dimension $n=2,3,4,6$. As already mentioned, Calabi-Yau manifolds are Ricci-flat and K\"ahler (at least to leading order). This means there is a real two form $J$, the K\"ahler form, which is closed. Some corrections may allow for a non-vanishing $dJ$, so the internal geometry is no longer K\"ahler and yet preserves supersymmetry. We discuss this possibility in detail for $n=4$.
In addition there is a holomorphic complex form $\Omega$ which, as opposed to the leading order Calabi-Yau case, may no longer be closed. Hence the term $SU(N)$-structure manifolds, as opposed to $SU(N)$-holonomy manifolds.

Perturbative corrections to Calabi-Yau compactifications of string theory were considered in the literature many years ago. See \cite{Witten:1985bz}, \cite{Gross:1986iv}, and \cite{Candelas:1986tz} for a partial list of references considering the effective action approach. A detailed analysis of the effect that these corrections have on a Calabi-Yau background was performed in the sigma model by \cite{Nemeschansky:1986yx} (see also \cite{Howe:1987qv} and \cite{Howe:1992tg}).
The novelty of our work, is that it makes contact to the more recent literature about string theory compactification on $SU(N)$-structure
manifolds (see \cite{Gauntlett:2002sc}, \cite{Martelli:2003ki}, \cite{Grana:2004bg}, \cite{KashaniPoor:2006si}, \cite{Fidanza:2003zi}, \cite{Stelle:2006ku} and \cite{Lu:2005im} for a partial list of references, and see especially \cite{Coimbra:2014qaa} for another work which studies string corrections in the language of $G$-structures, though in a somewhat different context).
The approach taken here is closely related to the effective field theory approach for M-theory and type II theory compactifications on $G_2$-structure manifolds and $\Spin(7)$-structure manifolds performed recently in \cite{Becker:2014rea}.  In this case, a conventional non-linear sigma model approach is still not available \cite{Nemeschansky:1986yx}.

Finally, we note that the one dimensional $(n=1)$ case is too simple because spinors have only one real component. Thus a unit normalized spinor just has a constant component whose absolute value is equal to one. There are no corrections allowed and the spinor is covariantly constant. The spin connection vanishes automatically in one dimension and the space (we assume compactness) is always $S^1$.  So the first case which isn't trivial is two dimensions. This is the case we work out next.

\section{Two dimensions}

Our goal in this section is to show that $T^2$ is the only compact internal two-dimensional geometry allowed by supersymmetry.
Even though such a proof seems fairly straightforward, the explicit construction of the G-structure is very instructive.
In two Euclidean dimensions we can pick gamma matrices $\G_a$ which are real and symmetric, and we can define the chirality operator
\be
\G=\frac{i}{2!}\e^{ab}\G_{ab},
\ee
which is imaginary, antisymmetric and squares to one.  Here $\e^{ab}$ is the antisymmetric tensor with entries $\pm 1/\sqrt{g}$.

Suppose we have a nowhere vanishing real spinor $\eta$.  This reduces the structure group from $\Spin(2)$ to $\{1\}$ (the trivial group).  We are free to normalize the spinor so that $\eta^T\eta=1$ pointwise. Then, again pointwise, one choice of basis for the two-dimensional space of real spinors is $\{\eta,i\G\eta\}$ which obeys the completeness relation
\be
1=\eta\eta^T+\G\eta\eta^T\G.
\ee
We can define one-forms in terms of these basis elements
\be
\label{eq:2DuvDef}
u_a=\eta^T\G_a\eta,\qquad v_a=i\eta^T\G_a\G\eta,
\ee
which are related by a duality transformation
\be
\label{eq:2dOrientation}
v=-\ast u.
\ee
Using the completeness relation, one can show that $u_a$ and $v_a$ are zweibeine.
\be
\label{eq:2dMetric}
u_au_b+v_av_b=g_{ab}.
\ee
Here and in the following sections $g_{ab}$ denotes the metric on $M$.
One can show (by contracting (\ref{eq:2dMetric}) in various ways), that the zeibeine satisfy
\be
\label{eq:2DContractionIdentities}
u^au_a=v^av_a=1,\qquad u^av_a=0.
\ee
We can either proceed by working with the metric and the spinor, or equivalently use an approach in terms of the zweibeine.
In fact, specifying a pair of globally defined real one-forms $u_a$ and $v_a$ such that $u\w v$ is nowhere vanishing is (almost\footnote{There are two caveats here if we want to be completely precise.  First of all as explained below, there's also the choice of orientation to be made.  Secondly, one can see that sending $\eta\rr -\eta$ leaves $u$ and $v$ unchanged, so there's an additional $\Z_2$ covering issue.  Neither of these subtleties will cause us problems, but it is good to keep them in mind.}) equivalent to specifying a metric and a unit normalized spinor $\eta$.  In terms of counting components, we have either three components from $g_{ab}$ and one from $\eta$, or we have two each from $u_a$ and $v_a$.  In either case the total number (four) parameterizes the space of $G$-structures ($G=\{1\}$ here), which indeed should have dimension
\be
\operatorname{dim}(\operatorname{GL}(2))-\operatorname{dim}(\{1\})=4-0=4.
\ee

The map between this data in one direction is given by (\ref{eq:2DuvDef}), because the spinor and the gamma matrices (alias the metric) determine the zweibeine.  For the other direction we can define a metric by (\ref{eq:2dMetric}).  Explicitly, the inverse metric is given by
\be
g^{-1}=\frac{1}{\lp u_1v_2-u_2v_1\rp^2}\lp\begin{matrix}(u_2)^2+(v_2)^2 & -u_1u_2-v_1v_2 \\ -u_1u_2-v_1v_2 & (u_1)^2+(v_1)^2\end{matrix}\rp.
\ee
From this expression one can directly show the various contraction identities (\ref{eq:2DContractionIdentities}).  Further, an appropriate choice of orientation (we need to pick the epsilon tensor to satisfy $\e_{ab}=u_av_b-v_au_b$) will ensure that (\ref{eq:2dOrientation}) holds as well.

What about the spinor $\eta$?  If we pick a choice of flat gamma matrices, say for instance
\be
\label{eq:2DFlatGammas}
\G_{\ul{u}}=\lp\begin{matrix} 1 & 0 \\ 0 & -1\end{matrix}\rp,\quad\G_{\ul{v}}=\lp\begin{matrix} 0 & 1 \\ 1 & 0\end{matrix}\rp,
\ee
(where an underline refers to a flat index), then the curved gamma matrices are expressed in terms of the zweibeine
\be
\G_a=u_a\G_{\ul{u}}+v_a\G_{\ul{v}}=\lp\begin{matrix} u_a & v_a \\ v_a & -u_a\end{matrix}\rp.
\ee
The explicit form of the chirality operator is
\be
\G=\lp\begin{matrix} 0 & i \\ -i & 0\end{matrix}\rp.
\ee
Now we need to pick a spinor $\eta$ that satisfies (\ref{eq:2DuvDef}).  It is easy to check that the solution is $\eta=(\begin{smallmatrix} 1 \\ 0\end{smallmatrix})$ (or $\eta=(\begin{smallmatrix} -1 \\ 0\end{smallmatrix})$).
This seems surprising at first, since there should have been one free component of $\eta$ to match our counting.  Looking more carefully, we note that there is one component of $u_a$ and $v_a$ which does not enter $g_{ab}$, and which corresponds to the freedom to make local $\SO(2)$ rotations.  When we make such a rotation we are essentially changing the choice of flat gamma matrices (\ref{eq:2DFlatGammas}), and this choice gets translated immediately into the missing component of $\eta$.  Explicitly, if we take
\be
\G_{\ul{u}}'=\cos\phi\G_{\ul{u}}+\sin\phi\G_{\ul{v}},\qquad\G_{\ul{v}}'=-\sin\phi\G_{\ul{u}}+\cos\phi\G_{\ul{v}},
\ee
with $\phi$ being some function on $M$, then the corresponding $\eta'$ would be given by
\be
\eta'=\lp\begin{matrix}\cos(\phi/2) \\ \sin(\phi/2)\end{matrix}\rp.
\ee
This shows the equivalence between the variables $\{g_{ab},\eta\}$ and $\{u_a,v_a\}$.

%
Working in terms of the metric/spinor pair, we can make the following ansatz for the corrected SUSY variation of the gravitino
\be
\label{eq:2dCorrectedVar}
\nabla_a\eta=A_a\eta+iB_a\G\eta.
\ee
Here we have used that any spinor can be expanded into our basis (\ref{eq:2DuvDef}), with coefficients $A_a$, $B_a$ that serve as source terms.
It is easy to see though, that $A_a=0$ for a spinor with unit norm (since $0=\nabla_a(\eta^T\eta)=2A_a$) and that $A_a=B_a=0$ for the vacuum.
Let's understand how to solve this equation for the metric/spinor pair and then we'll equivalently think about the question in terms of the one-forms $u$ and $v$. The integrability condition obtained by hitting (\ref{eq:2dCorrectedVar}) by another nabla and antisymmetrizing takes the form
\be
\label{eq:2dIntegrability}
\nabla_{[a}B_{b]}=-\frac{1}{8}\e^{cd}R_{abcd}=-\frac{R}{8}\e_{ab},
\ee
where we used the fact that in two dimensions
\be
R_{abcd}=\frac{R}{2}\lp g_{ac}g_{bd}-g_{ad}g_{bc}\rp.
\ee
Since $B_a$ encodes the (perturbative and possibly non-perturbative corrections), it has an expansion in derivatives, vanishing at lowest order.  Then at lowest order (\ref{eq:2dIntegrability}) implies that $R_{abcd}=0$, so the space is (assuming also compactness and orientability) simply $T^2$ with a flat metric.  We are free to choose coordinates in which the metric is simply constant.  In these coordinates the spin connection vanishes and so we see that $\eta$ must also be constant.  At the next order (and every higher order), $B_a$ will be some local covariant construction in terms of curvatures and derivatives, which vanishes when evaluated on $T^2$. Thus $B_a=0$ to all orders in perturbation theory
(and even non-perturbatively) and the solution remains a flat torus. For the same reason the spinor is constant to all orders.

How do we see this if we translate the problem to $\{u,v\}$?  By taking derivatives of (\ref{eq:2DuvDef}) we obtain
\be
\nabla_au_b=2B_av_b,\qquad\nabla_av_b=-2B_au_b.
\ee
As it is these equations still depend on the metric through the covariant derivatives.  But by antisymmetrizing, we obtain
\be
\label{eq:2DGStructEqns}
du=2B\w v,\qquad dv=-2B\w u.
\ee
If we now view $u_a$ and $v_a$ as independent quantities, and $B_a$ is a functional of $u_a$ and $v_a$ (via the metric $g_{ab}[u,v]$ as an intermediary), then (\ref{eq:2DGStructEqns}) are equations that we need to solve.  In fact it will turn out that these equations are equivalent to (\ref{eq:2dCorrectedVar}), as will become clear by our solution.

Recalling the identification of $u$ and $v$ as zweibeine, (\ref{eq:2DGStructEqns}) identifies $B_a$ as the spin connection,
\be
B_a=-\hlf\lp\om_a\rp^{\ul{u}}_{\hph{\ul{u}}\ul{v}}.
\ee
Both sides of this equation should be viewed as functionals of $u$ and $v$, and the equation itself is a constraint on $u$ and $v$.  At leading order $B_a=0$, so we have vanishing spin connection and hence our manifold is $T^2$ with a flat metric. These statements hold to all orders in the perturbative expansion(and even non-perturbatively) because $B_a=0$, when evaluated in the flat background solution. We can then choose $u$ and $v$ to be the zweibeine compatible with these statements.  Note that given such a choice of $u$ and $v$, we can always add exact one-forms and still have a solution (at least if the deformation is small enough to not spoil the condition $u\w v\ne 0$).  This corresponds to performing a diffeomorphism, and we shall see an analog to this freedom in each dimension.

%

\section{Three dimensions}

We will be somewhat briefer in this case, because the arguments work along the same lines as in two dimensions. In this case $T^3$ is the only compact manifold allowed by supersymmetry, as we discuss next.
In three Euclidean dimensions, the two-by-two gamma matrices $\G_a$ cannot be chosen to be real or have particular symmetry properties (other than being hermitian). There is a unitary charge conjugation matrix $C$ which is antisymmetric and satisfies $\G_a^T=-C^{-1}\G_aC$.  For example, if we take the standard Pauli matrices $\s_a$ for $\G_a$, then we can take $C=\s_2$.  There is a relation
\be
\frac{i}{3!}\e^{abc}\G_{abc}=1.
\ee

Suppose we have a nowhere vanishing complex spinor $\eta$.  This reduces the structure group from $\Spin(3)$ to $\{1\}$.  We can normalize $\eta^\dagger\eta=1$, which still leaves an unfixed phase of $\eta$.  A complex basis for the spinors is then given by $\eta$ and $C\bar{\eta}$ (the bar denoting complex conjugation), the latter also being normalized $(C\bar{\eta})^\dagger(C\bar{\eta})=1$, since $C$ is unitary.  We also have orthogonality, $\eta^\dagger(C\bar{\eta})=0$, by the antisymmetry of $C$.  The corresponding completeness relation satisfied by these basis elements is
\be
\label{eq:3dCompleteness}
1=\eta\eta^\dagger+C\bar{\eta}\lp C\bar{\eta}\rp^\dagger.
\ee

In terms of this basis there are three real one forms we can construct
\be
u_a=\eta^\dagger\G_a\eta,
\ee
and
\be
v_a=\hlf\ls\lp C\bar{\eta}\rp^\dagger\G_a\eta+\eta^\dagger\G_a\lp C\bar{\eta}\rp\rs,\quad w_a=\frac{i}{2}\ls\lp C\bar{\eta}\rp^\dagger\G_a\eta-\eta^\dagger\G_a\lp C\bar{\eta}\rp\rs.
\ee
The remaining one-form is not independent, since
\be
\lp C\bar{\eta}\rp^\dagger\G_a\lp C\bar{\eta}\rp=\eta^TC^{-1}\G_aC\bar{\eta}=-\eta^T\G_a^T\bar{\eta}=-u_a.
\ee

This triplet of one-forms are an equivalent parametrization of the data of the $G$-structure.  Indeed, we can check that counting matches, since
\be
(\mathrm{\#\ of\ real\ components\ of\ }g\ \mathrm{and}\ \eta)=6+3=9,
\ee
\be
(\mathrm{\#\ of\ components\ of\ }\{u,v,w\})=3\times 3=9,
\ee
and the space of $G$-structures has dimension
\be
\operatorname{dim}(\operatorname{GL}(3))-\operatorname{dim}(\{1\})=9-0=9.
\ee

The one-forms behave as a set of dreibeine, since we can show using (\ref{eq:3dCompleteness}) that
\be
u_au_b+v_av_b+w_aw_b=g_{ab}.
\ee
Working with the metric/spinor pair is again equivalent to working with the dreibeine defined above.
In the former approach, we write the susy transformation of the gravitino in terms of our basis of spinors
\be
\nabla_a\eta=iA_a\eta+B_aC\bar{\eta}.
\ee
Since we should preserve the normalization $\eta^\dagger\eta=1$, it follows that $A_a$ here should be real, while $B_a$ may be complex.  In terms of the dreibeine introduced above, $A_a$ and $B_a$ are the nine real components of the spin-connection. Since $A_a$ and $B_a$ vanish to lowest order, the unique compact solution is a flat torus $T^3$ and a constant complex spinor. By the same arguments as in two dimensions, $A_a$ and $B_a$ vanish to all orders when evaluated on the background $T^3$. So a flat three dimensional torus is the unique solution to all orders in perturbation theory and non-perturbatively. Similarly, the spinor is constant to all orders.
We observe that to leading order the dreibeine, $u$, $v$, and $w$ are all Killing vectors and the associated metric is flat. By the same argument that we used in two dimensions, this holds to all orders.  As in two dimensions, infinitessimal diffeomorphisms are incorporated by the freedom to add arbitrary exact one-forms to $u$, $v$, and $w$.

\section{Four dimensions}
In this dimension, corrections can change the geometry in a more interesting manner. Supersymmetry allows for two compact manifolds, as we discuss next.

\subsection{Spinors and bispinors}
\label{subsec:4dBispinors}

In four dimensions, there is again no basis in which the $4\times 4$ $\G_a$ can be chosen to be real or pure imaginary.  We define the chirality operator
\be
\G=\frac{1}{4!}\e^{abcd}\G_{abcd},
\ee
which satisfies $\G^2=1$.  In fact we can choose a basis in which $\G$ is real and symmetric.  Note also that
\be
\label{eq:4dCliffordHodge}
\G\G_{ab}=-\hlf\e_{abcd}\G^{cd}.
\ee

The unitary charge conjugation matrix $C$ is antisymmetric, satisfies\footnote{We can actually choose either sign in this relation, $\G_a^T=\pm C\G_aC^{-1}$; there exist conjugation matrices for both.} $\G_a^T=-C^{-1}\G_aC$ and commutes with $\G$.

Suppose we have a nowhere vanishing complex spinor $\eta$ with a particular chirality, say\footnote{This choice ensures that $J$ and $\Om$ are self-dual, which is standard in the literature, rather than anti-self-dual.} $\G\eta=-\eta$, which we can normalize so that $\eta^\dagger\eta=1$.  This reduces the structure group from $\Spin(4)$ to $\SU(2)$.  Then, analogous to three dimensions, $\eta$ and $C\bar{\eta}$ build a complex basis for the positive-chirality spinors, and we have a corresponding completeness relation
\be
\label{eq:4dCompleteness}
\frac{1-\G}{2}=\eta\eta^\dagger+C\bar{\eta}\lp C\bar{\eta}\rp^\dagger.
\ee

There are no real one-forms that we can construct in this case, but there are three real two-forms, or one real and one complex two-form,
\be
J_{ab}=i\eta^\dagger\G_{ab}\eta,\qquad\Om_{ab}=\eta^\dagger\G_{ab}C\bar{\eta}.
\ee
The remaining possibility is not independent, since
\be
i\lp C\bar{\eta}\rp^\dagger\G_{ab}C\bar{\eta}=i\eta^TC^{-1}\G_{ab}C\bar{\eta}=i\eta^\dagger C\lp\G_{ab}\rp^TC^{-1}\eta=-J_{ab}.
\ee
Using (\ref{eq:4dCliffordHodge}), we can show that these forms are self-dual  $\ast J=J$ and $\ast\Om=\Om$.
By expanding the combinations $\eta\eta^\dagger$ in terms of the even elements of the Clifford algebra (i.e.\ in terms of products of even numbers of gamma matrices), we can refine (\ref{eq:4dCompleteness}), obtaining Fierz identities
\be
\eta\eta^\dagger=\hlf\frac{1-\G}{2}+\frac{i}{8}J^{ab}\G_{ab},\qquad C\bar{\eta}\lp C\bar{\eta}\rp^\dagger=\hlf\frac{1-\G}{2}-\frac{i}{8}J^{ab}\G_{ab},
\ee
and
\be
C\bar{\eta}\eta^\dagger=-\frac{1}{8}\Om^{ab}\G_{ab},\qquad\eta\lp C\bar{\eta}\rp^\dagger=\frac{1}{8}\overline{\Om}^{ab}\G_{ab}.
\ee
From these we can derive contraction identities, which naturally will involve the metric
\be
J^{ac}J_{bc}=\d^a_b,\quad J^{ac}\Om_{bc}=-i\Om^a_{\hph{a}b},\quad\Om^{ac}\Om_{bc}=0,\quad \Om^{ac}\overline{\Om}_{bc}=2\d^a_b-2iJ^a_{\hph{a}b},
\ee
and their complex conjugates. From this further contractions can be derived
\be
J^{ab}J_{ab}=4,\quad J^{ab}\Om_{ab}=0,\quad\Om^{ab}\Om_{ab}=0,\quad\Om^{ab}\overline{\Om}_{ab}=8.
\ee
We can also show that there are metric independent constraints relating the various two forms
\be
J\w\Om=\Om\w\Om=0,\qquad J\w J=\hlf\Om\w\overline{\Om}\ne 0.
\ee
Equivalently, we can reformulate everything with the replacement of the complex two-form $\Om$ by two real two-forms,
\be
\Om_{1\,ab}=\hlf\lp\Om_{ab}+\overline{\Om}_{ab}\rp,\qquad\Om_{2\,ab}=-\frac{i}{2}\lp\Om_{ab}-\overline{\Om}_{ab}\rp.
\ee
The contraction identities then become
\be
J^{ac}\Om_{1\,bc}=\Om_{2\hph{a}b}^{\hph{2}a},\qquad J^{ac}\Om_{2\,bc}=-\Om_{1\hph{a}b}^{\hph{1}a},\non
\ee
\be
\Om_1^{ac}\Om_{1\,bc}=\Om_2^{ac}\Om_{2\,bc}=\d^a_b,\qquad\Om_1^{ac}\Om_{2\,bc}=J^a_{\hph{a}b},
\ee
while the metric independent constraints read
\be
\label{eq:4dFormConstraints}
J\w\Om_1=J\w\Om_2=\Om_1\w\Om_2=0,\qquad J\w J=\Om_1\w\Om_1=\Om_2\w\Om_2.
\ee
The triplet of real two-forms $\{J,\Om_1,\Om_2\}$ plus the constraints (\ref{eq:4dFormConstraints}) are equivalent data to the original $\{g_{ab},\eta\}$.  As a check, we have the counting
\be
{(\mathrm{\#\ of\ components\ of\ }\{g,\eta\})=10+3=13,}
\ee
\begin{multline}
(\mathrm{\#\ of\ components\ of\ }\{J,\Om_1,\Om_2\})-(\mathrm{\#\ of\ constraints})=3\times 6-5\\
=13,
\end{multline}
and
\be
\operatorname{dim}(\operatorname{GL}(4))-\operatorname{dim}(\SU(2))=16-3=13.
\ee
The forms $J,\Om_1,\Om_2$ satisfying the constraints  (\ref{eq:4dFormConstraints}) build an $SU(2)$-structure and the four-dimensional manifold is called an $SU(2)$-structure manifold.
These manifolds include, in particular, $SU(2)$-holonomy manifolds, for which all three two forms are closed.
Such manifolds are also called hyperK\"ahler.
Having an $SU(2)$-structure is equivalent to having a metric plus a spinor of constant norm. This equivalence is again simple in
one direction. Namely, having a metric and a normalized spinor one can define two forms by (4.4), that satisfy the constraints (4.13).
In turn, having an $SU(2)$-structure defines a metric and a normalized spinor. In the next subsection we illustrate how to compute the metric in terms of $J$, $\Om_1$, and $\Om_2$.

\subsection{Metric from $\SU(2)$-structure}
\label{subsec:4DMetric}

Suppose we are given a triplet of real two-forms $J$, $\Om_1$, and $\Om_2$, which satisfy (\ref{eq:4dFormConstraints}). Using the antisymmetric symbol (not tensor) $\hat{\e}$, we can define
\be
s_{ab}=\hlf\hat{\e}^{cdef}\Om_{1\,(a|c|}\Om_{2\,b)d}J_{ef}.
\ee
Let $s=\det(s)$, and as long as $s\ne 0$, define the metric in terms of $s_{ab}$
\be
g_{ab}=s^{-1/6}s_{ab}.
\ee
With this metric and its inverse, one can verify all the contraction identities of section \ref{subsec:4dBispinors}.

%

\subsection{Decomposition of forms}
In this section we present some useful statements about the decomposition of forms on $SU(2)$-structure manifolds that are needed in the following subsection.
Under $\SO(4)\rr\SU(2)$, the spaces $\La^p$ of differential $p$-forms (at some given point on the manifold) decompose as
\bea
\La^0 &\cong& \La^0_1,\\
\La^1 &\cong& \La^1_{2\oplus 2},\\
\La^2 &\cong& \La^2_{1\oplus 1\oplus 1}\oplus\La^2_3,\\
\La^3 &\cong& \La^3_{2\oplus 2},\\
\La^4 &\cong& \La^4_0,
\eea
where the subscripts on the right-hand-side indicate the $\SU(2)$ representations.  For the one-forms (and also the three-forms), the individual representations $\mathbf{2}$ are not real representations (they are pseudo real), which is why we combine them in one subscript.  They can be thought of as the $\pm i$ eigenspaces of the matrix $J_a^{\hph{a}b}$, which squares to $-1$.  For the two-forms, the three singlet representations correspond simply to the real forms $J$, $\Om_1$, and $\Om_2$.
Note that all of these singlet two-forms are self-dual, as we saw in section \ref{subsec:4dBispinors}.  The total space of two-forms should split into anti-self-dual and self-dual subspaces,
\be
\La^2\cong\La^2_{ASD}\oplus\La^2_{SD},
\ee
and the two spaces should have the same dimension. Remember we are talking about spaces of forms at a point, not about cohomologies, where the dimensions of these spaces are typically not equal). The fact that $\La^2_{1\oplus 1\oplus 1}\subset\La^2_{SD}$ implies
\be
\La^2_{SD}\cong\La^2_{1\oplus 1\oplus 1},\qquad\La^2_{ASD}\cong\La^2_3.
\ee

Using this information, let's take a brief detour to prove a useful statement that is needed in the following. Suppose we are on a compact oriented $4k$ dimensional Riemannian manifold.  For a $(2k-1)$-form $\xi$, the Hodge decomposition is
\be
\xi=d\al+d^\dagger\beta+\g,
\ee
where $\al$ is a $(2k-2)$-form, $\beta$ is a $(2k)$-form, and $\g$ is a harmonic $(2k-1)$-form.  We claim that we can in fact take $\beta$ to be self-dual.  Indeed, if $\beta$ is not self-dual, then take its Hodge decomposition,
\be
\beta=d\m+d^\dagger\n+\om,
\ee
and let
\be
\beta'=d\m+\ast d\m=d\m+d^\dagger\lp\ast\m\rp.
\ee
Then $\beta'$ is manifestly self-dual and we also have $d^\dagger\beta'=d^\dagger\beta$, so we can write
\be
\xi=d\al+d^\dagger\beta'+\g.
\ee
But returning now to our particular case, any self-dual two-form can be expanded in terms of $J$, $\Om_1$ and $\Om_2$
\be
\beta'=xJ+y\Om_1+z\Om_2.
\ee
Thus, on our $\SU(2)$-structure manifold, any one-form can be written as
\be
\label{eq:OneFormDecomp}
\xi=d\la+d^\dagger\lp xJ+y\Om_1+z\Om_2\rp+\g,
\ee
where $\la$, $x$, $y$, and $z$ are functions.  In components,
\be
\label{eq:OneFormComponents}
\xi_a=\nabla_a\la+\nabla^b\lp xJ_{ab}+y\Om_{1\,ab}+z\Om_{2\,ab}\rp+\g_a.
\ee

\subsection{Gravitino SUSY variation and background geometry}

Additional information about which $SU(2)$-structure manifolds are relevant for M-theory/string theory follows from the gravitino supersymmetry transformation, which involves our basis of spinors
\be
\label{eq:4DSpinorVariation}
\nabla_a\eta=iA_a\eta+B_aC\bar{\eta}.
\ee
We will not need concrete expressions for the source terms $A_a$ and $B_a$, which depend on the concrete theory one wishes to consider. The above is the most general expression allowed by supersymmetry. For example, if perturbative corrections to the internal geometry are allowed, these can be calculated from scattering amplitudes, and would contribute to $A_a$ and/or $B_a$. These may involve Riemann tensors, covariant derivatives, and perhaps also the forms $J$ and $\Om_i$. Fluxes and non-perturbative effects may also be allowed in certain theories. If additional fields are being considered, their SUSY constraints need to be checked though.
Note, however, that $A_a$ is real in order to preserve the normalization of $\eta$, but $B_a$ may be complex, $B_a=B_{1\,a}+iB_{2\,a}$.  The twelve real components of the one-forms $A_a$ and $B_a$ match onto the components of the three unconstrained real three-forms $dJ$, $d\Om_1$, and $d\Om_2$.  Indeed, we have
\be
\label{eq:4DGeneralFormVariation}
dJ=-2B_1\w\Om_2-2B_2\w\Om_1,\ d\Om_1=2A\w\Om_2+2B_2\w J,\ d\Om_2=-2A\w\Om_1+2B_1\w J.
\ee
$A_a$ and $B_a$ vanish to leading order, so to leading order $dJ=d\Om=0$.  The metric $g_{ab}$ is thus Calabi-Yau (K\"ahler and Ricci flat). In fact, these manifolds are actually hyperK\"ahler. We know that if the space is compact it is either $T^4$ or K3.  In the case of $T^4$, there are no higher curvature invariants; $A_a$ and $B_a$ remain zero to all orders and the solution is uncorrected. The arguments follow along the same lines as in the lower dimensional cases.

Suppose instead, that the leading order space is K3, where corrections might be allowed. See e.g \cite{Howe:1992tg} for a discussion of perturbative corrections in the context of $(0,4)$ sigma models or \cite{Strominger:1986uh} for an example involving fluxes.

In the following we use primes to denote corrected quantities, and unprimed objects will correspond to the underlying K3.  The integrability condition implies that the right-hand sides of (\ref{eq:4DGeneralFormVariation}) are closed, i.e.\ that
\be
0=-dB_1\w\Om_2-dB_2\w\Om_1=dA\w\Om_2+dB_2\w J=-dA\w\Om_1+dB_1\w J.
\ee
Note that this allows for the internal geometry to be non-K\"ahler, i.e. $dJ$ to be non-vanishing.
Let's understand the integrability constraints better.  By contracting with the volume form and using the self-duality of the forms $J$, $\Om_1$, and $\Om_2$, we can recast these equations as
\be
0=\Om_2^{ab}\nabla_aB_{1\,b}+\Om_1^{ab}\nabla_aB_{2\,b}=\Om_2^{ab}\nabla_aA_b+J^{ab}\nabla_aB_{2\,b}=\Om_1^{ab}\nabla_aA_b-J^{ab}\nabla_aB_{1\,b}.
\ee
We now use (\ref{eq:OneFormComponents}) and the fact that the leading order $J$, $\Om_1$, and $\Om_2$ are all covariantly constant and that there are no harmonic one-forms on K3, to write
\bea
A_a &=& \nabla_a\lambda_A+J_a^{\hph{a}b}\nabla_bx_A+\Om_{1\,a}^{\hph{1\,a}b}\nabla_by_A+\Om_{2\,a}^{\hph{2\,a}b}\nabla_bz_A,\\
B_{1\,a} &=& \nabla_a\lambda_{B_1}+J_a^{\hph{a}b}\nabla_bx_{B_1}+\Om_{1\,a}^{\hph{1\,a}b}\nabla_by_{B_1}+\Om_{2\,a}^{\hph{2\,a}b}\nabla_bz_{B_1},\\
B_{2\,a} &=& \nabla_a\lambda_{B_2}+J_a^{\hph{a}b}\nabla_bx_{B_2}+\Om_{1\,a}^{\hph{1\,a}b}\nabla_by_{B_2}+\Om_{2\,a}^{\hph{2\,a}b}\nabla_bz_{B_2}.
\eea
In terms of these components, the integrability conditions become simply
\be
0=\nabla^2\lp z_{B_1}+y_{B_2}\rp=\nabla^2\lp z_A+x_{B_2}\rp=\nabla^2\lp y_A-x_{B_1}\rp.
\ee
Note that we are free to shift any of the $\la$'s, $x$'s, $y$'s, or $z$'s by constants, since this will not change the one-forms.  Hence, since constants are also the only harmonic functions on a compact connected space, the fully general solution to the integrability conditions is
\be
0=z_{B_1}+y_{B_2}=z_A+x_{B_2}=y_A-x_{B_1}.
\ee

Now put these decompositions into the equation (\ref{eq:4DGeneralFormVariation})
\bea
\lp dJ'\rp_{abc} &=& -6B_{1\,[a}\Om_{2\,bc]}-6B_{2\,[a}\Om_{1\,bc]}\\
&=& -6\lp\nabla_{[a}\la_{B_1}+J_{[a}^{\hph{[a}d}\nabla_{|d|}x_{B_1}+\Om_{1\,[a}^{\hph{1\,[a}d}\nabla_{|d|}y_{B_1}+\Om_{2\,[a}^{\hph{2\,[a}d}\nabla_{|d|}z_{B_1}\rp\Om_{2\,bc]}\non\\
&& -6\lp\nabla_{[a}\la_{B_2}+J_{[a}^{\hph{[a}d}\nabla_{|d|}x_{B_2}+\Om_{1\,[a}^{\hph{1\,[a}d}\nabla_{|d|}y_{B_2}+\Om_{2\,[a}^{\hph{2\,[a}d}\nabla_{|d|}z_{B_2}\rp\Om_{1\,bc]}.\non
\eea
To proceed further, note that we have identities such as
\be
J_{[a}^{\hph{[a}d}\nabla_{|d|}x\Om_{2\,bc]}=-\nabla_{[a}x\Om_{1\,bc]},
\ee
and others related by permutation of the three-forms, which can be derived using the self-duality of the forms and the contraction identities of section \ref{subsec:4dBispinors}.  From this one can show that the equation above becomes in fact
\be
dJ'=2d\ls\lp -y_{B_1}+z_{B_2}\rp J+\lp x_{B_1}-\la_{B_2}\rp\Om_1+\lp -\la_{B_1}-x_{B_2}\rp\Om_2\rs.
\ee
Note that to reach this form we must make use of the fact $z_{B_1}+y_{B_2}=0$ which we derived from integrability.

Thus, we should make an ansatz for the $SU(2)$ structure
\be
J'=J+2\lp -y_{B_1}+z_{B_2}\rp J+2\lp x_{B_1}-\la_{B_2}\rp\Om_1+2\lp -\la_{B_1}-x_{B_2}\rp\Om_2+da,
\ee
where $a$ is some arbitrary one-form.

Similarly with the other forms, we find that we should write
\be
\Om_1'=\Om_1+2\lp y_A+\la_{B_2}\rp J+2\lp -x_A+z_{B_2}\rp\Om_1+2\lp\la_A-y_{B_2}\rp\Om_2+db,
\ee
\be
\Om_2'=\Om_2+2\lp z_A+\la_{B_1}\rp J+2\lp -\la_A+z_{B_1}\rp\Om_1+2\lp -x_A-y_{B_1}\rp\Om_2+dc,
\ee
where again $b$ and $c$ are one-forms.  These expressions for $J'$, $\Om_1'$, and $\Om_2'$ now automatically solve the variation equations (\ref{eq:4DGeneralFormVariation}), but we still have to do some work to ensure that the constraints (\ref{eq:4dFormConstraints}).  These will constrain the one-forms $a$, $b$, and $c$.

To tackle this next step, let us use (\ref{eq:OneFormComponents}) again to expand $a$, $b$, and $c$,
\bea
a &=& \nabla_a\la_1+J_a^{\hph{a}b}\nabla_bx_1+\Om_{1\,a}^{\hph{1\,a}b}\nabla_by_1+\Om_{2\,a}^{\hph{2\,a}b}\nabla_bz_1,\\
b &=& \nabla_a\la_2+J_a^{\hph{a}b}\nabla_bx_2+\Om_{1\,a}^{\hph{1\,a}b}\nabla_by_2+\Om_{2\,a}^{\hph{2\,a}b}\nabla_bz_2,\\
c &=& \nabla_a\la_3+J_a^{\hph{a}b}\nabla_bx_3+\Om_{1\,a}^{\hph{1\,a}b}\nabla_by_3+\Om_{2\,a}^{\hph{2\,a}b}\nabla_bz_3.
\eea
In terms of these, the orthogonality constraints become
\bea
0 &=& \nabla^2\lp y_1+x_2\rp-4\lp y_A+x_{B_1}\rp,\\
0 &=& \nabla^2\lp z_1+x_3\rp-4\lp z_A-x_{B_2}\rp,\\
0 &=& \nabla^2\lp z_2+y_3\rp-4\lp z_{B_1}-y_{B_2}\rp,
\eea
while the normalization conditions lead to
\be
\nabla^2x_1-4\lp -y_{B_1}+z_{B_2}\rp=\nabla^2y_2-4\lp -x_A+z_{B_2}\rp=\nabla^2z_3-4\lp -x_A-y_{B_1}\rp.
\ee
These give five Poisson equations for the nine functions $x_i$, $y_i$, and $z_i$, where $i$ runs from one to three.  Note that the $\la_i$ of course drop out, since they don't appear in $da$, $db$, or $dc$.  Furthermore, these Poisson equations can always be solved, once we make use of the freedom to shift the source terms by constants.

To summarize, we can pick four of the functions, say $y_1$, $z_1$, $z_2$, and $y_2$, arbitrarily, and then fix the remaining functions by solving Poisson equations,
\bea
\nabla^2x_1 &=& \nabla^2y_2+4\lp x_A-y_{B_1}\rp,\\
\nabla^2x_2 &=& -\nabla^2y_1+4\lp y_A+x_{B_1}\rp,\\
\nabla^2x_3 &=& -\nabla^2z_1+4\lp z_A-x_{B_2}\rp,\\
\nabla^2y_3 &=& -\nabla^2z_2+4\lp z_{B_1}-y_{B_2}\rp,\\
\nabla^2z_3 &=& \nabla^2y_2+4\lp -y_{B_1}-z_{B_2}\rp.
\eea
Here all quantities labeled with subindices $A$, $B$ need external input and act as source terms, coming from string theory. Their explicit form is not relevant for our discussion.
Once this is done, then the corrected $\SU(2)$-structure, or equivalently the corrected metric and spinor, solve the full system of equations.  Since we can always do this, it means that we can always correct the K3 metric in order to preserve SUSY to this order.  The new physical metric is no longer Ricci-flat nor K\"ahler (for general $A_a$ and $B_a$).

Let's understand this solution in a bit more detail.  First of all, what is the significance of the four arbitrary functions?  To shed light on their role, let us compute the correction to the metric.  By taking variations of the contraction identity
\be
g^{cd}\Om_{1\,ac}\Om_{2\,bd}=J_{ab},
\ee
we derive
\be
-\Om_{1\,a}^{\hph{1\,a}c}\Om_{2\,b}^{\hph{2\,b}d}\d g_{cd}+\Om_{2\,b}^{\hph{2\,b}c}\d\Om_{1\,ac}+\Om_{1\,a}^{\hph{1\,a}c}\d\Om_{2\,bc}=\d J_{ab},
\ee
which we can rearrange to get
\be
\d g_{ab}=\Om_{1\,a}^{\hph{1\,a}c}\d\Om_{1\,bc}+\Om_{2\,b}^{\hph{2\,b}c}\d\Om_{2\,ac}-\Om_{1\,a}^{\hph{1\,a}c}\Om_{2\,b}^{\hph{2\,b}d}\d J_{cd}.
\ee
Plugging in $J'$, $\Om_1'$, and $\Om_2'$,
\begin{multline}
\d g_{ab}=-4x_Ag_{ab}+\lp\Om_{1\,(a}^{\hph{1\,(a}c}\Om_{1\,b)}^{\hph{1\,b)}d}+\Om_{2\,(a}^{\hph{2\,(a}c}\Om_{2\,b)}^{\hph{2\,b)}d}\rp\nabla_c\nabla_d\lp x_1-y_2\rp\\
+\lp\d_{(a}^c\Om_{2\,b)}^{\hph{2\,b)}d}-J_{(a}^{\hph{(a}c}\Om_{1\,b)}^{\hph{1\,b)}d}\rp\nabla_c\nabla_d\lp x_2+y_1\rp\\
+\lp -\d_{(a}^c\Om_{1\,b)}^{\hph{1\,b)}d}-J_{(a}^{\hph{(a}c}\Om_{2\,b)}^{\hph{2\,b)}d}\rp\nabla_c\nabla_d\lp x_3+z_1\rp\\
+\lp\d_{(a}^cJ_{b)}^{\hph{b)}d}-\Om_{1\,(a}^{\hph{1\,(a}c}\Om_{2\,b)}^{\hph{2\,b)}d}\rp\nabla_c\nabla_d\lp y_3+z_2\rp\\
+\lp -\d_{(a}^c\d_{b)}^d-\Om_{2\,(a}^{\hph{2\,(a}c}\Om_{2\,b)}^{\hph{2\,b)}d}\rp\nabla_c\nabla_d\lp z_3-y_2\rp+\nabla_a\xi_b+\nabla_b\xi_a,
\end{multline}
where we have defined
\be
\xi_a=-\nabla_ay_2-J_a^{\hph{a}b}\nabla_bz_2+\Om_{1\,a}^{\hph{1\,a}b}\nabla_bz_1-\Om_{2\,a}^{\hph{2\,a}b}\nabla_by_1.
\ee
Comparing to (\ref{eq:OneFormComponents}), we see that by choosing $y_1$, $z_1$, $y_2$, and $z_2$ we can get the most general possible vector $\xi_a$, and that any choice of $\xi_a$ generates an infinitessimal diffeomorphism of the metric.  So the arbitrariness of the choices of these four functions simply corresponds to the possibility of arbitrary infinitessimal changes of coordinates.

Now let us briefly restrict to the situation when $B_a=0$.  In this case all the functions with $B_1$ or $B_2$ subscripts vanish, and the integrability conditions imply that $y_A=z_A=0$, leaving only $\la_A$ and $x_A$.  Taking $y_1=z_1=y_2=z_2=0$, the Poisson equations are solved by setting $x_2=x_2=y_3=z_3=0$, and solving
\be
\nabla^2x_1=4x_A.
\ee
The corrected $\SU(2)$-structure is in this simpler case
\be
J'=J+da,\quad\Om_1'=\Om_1+2\la_A\Om_2-2x_A\Om_1,\quad\Om_2'=\Om_2-2\la_A\Om_1-2x_A\Om_2,
\ee
where
\be
a_a=J_a^{\hph{a}b}\nabla_bx_1.
\ee
The corrected metric remains K\"ahler, $dJ'=0$, but $\Om_1'$ and $\Om_2'$ are no longer necessarily closed.  As mentioned before, the corrected physical metric is no longer Ricci-flat. Therefore the new manifold is not Calabi-Yau in the metric sense. It is nevertheless, Calabi-Yau in the topological sense, as it is K\"ahler and has $c_1(M)=0$.

The role of $x_1$ is somewhat clarified if we adopt local complex coordinates in which $J_i^{\hph{i}j}=i\d_i^j$, $J_{\bar{\imath}}^{\hph{\bar{\imath}}\bar{\jmath}}=-i\d_i^j$, and $J_i^{\hph{i}\bar{\jmath}}=J_{\bar{\imath}}^{\hph{\bar{\imath}}j}=0$.  Then if we define a complex scalar $\varphi=\la_A+ix_A$, we have $A_i=\partial_i\varphi$, $A_{\bar{\imath}}=\partial_{\bar{\imath}}\overline{\varphi}$.  The complex two-form becomes $\Om'=(1-2i\overline{\varphi})\Om$.  Finally, we have $a_i=i\partial_ix_1$, $a_{\bar{\imath}}=-i\partial_{\bar{\imath}}x_1$, and thus
\be
J'_{ij}=J'_{\bar{\imath}\bar{\jmath}}=0,\qquad J'_{i\bar{\jmath}}=J_{i\bar{\jmath}}-2i\partial_i\partial_{\bar{\jmath}}x_1.
\ee
In other words, $x_1$ is simply a shift in the K\"ahler potential.

Finally, returning to the general case, notice that $\la_A$, $\la_{B_1}$, and $\la_{B_2}$ appeared in a rather trivial way, dropping out of the Poisson equations.  In fact, we could have removed the $\la$'s by taking a different choice of nowhere vanishing positive chirality spinor.  Indeed, we can a priori make an $\SU(2)$ rotation in the space spanned by $\eta$ and $C\bar{\eta}$, to define a new spinor
\be
\hat\eta=\al\eta+\beta C\bar{\eta},\qquad\al,\beta\in\CC,\qquad\left|\al\right|^2+\left|\beta\right|^2=1.
\ee
If $\eta$ obeyed the SUSY variation (\ref{eq:4DSpinorVariation}), then $\hat{\eta}$ will obey a similar equation, but where $A_a$ and $B_a$ get modified.  Choosing appropriate infinitessimal rotations, we can in fact shift $A_a$ and $B_a$ by exact quantities, thus effectively shifting the $\la$'s.

\section{Six dimensions}
M-theory/string theory corrections, in particular perturbative effects coming from higher derivatives or $\al'$-corrections,
are possible upon compactification on a 6D K\"ahler manifold \cite{Gross:1986iv}, \cite{Nemeschansky:1986yx} and \cite{Candelas:1986tz}.
In this case the physical metric may no longer be Ricci flat, but nevertheless $c_1(M)=0$.
The corrected geometry has $SU(3)$-structure, rather than $SU(3)$-holonomy, as we show explicitly in the following. See \cite{Stelle:2006ku} and references therein for related discussions.

\subsection{Spinors and bispinors}

In six Euclidean dimensions we can pick gamma-matrices $\G_a$ which are imaginary and antisymmetric.  We can define
the chirality operator
\be
\G=\frac{i}{6!}\e^{abcdef}\G_{abcdef},
\ee
which is also imaginary antisymmetric.  Here $\e^{abcdef}$ is the antisymmetric tensor with entries $\pm 1/\sqrt{g}$.

Suppose we have a nowhere vanishing real spinor $\eta$.  This reduces the structure group from $\SO(6)$ to $\SU(3)$ which leaves the spinor invariant.  We are free to normalize the spinor so that $\eta^T\eta=1$ pointwise.  Then, again pointwise, one choice of basis for the eight dimensional space of real spinors is $\{\eta,i\G\eta,i\G_a\eta\}$.  The corresponding dual basis is given by $\{\eta^T,-i\eta^T\G,-i\eta^T\G^a\}$.  The completeness relation for this basis is the Fierz identity
\be
\label{eq:FirstFierz}
1=\eta\eta^T+\G\eta\eta^T\G+\G^a\eta\eta^T\G_a.
\ee
Along with the zero form $1$ and the six-form $\e_{abcdef}$, the existence of $\eta$ allows us to define the following real forms, which by construction are invariant under the $\SU(3)$-structure group
\be
J_{ab}=-i\eta^T\G_{ab}\G\eta,\qquad L_{abcd}=-\eta^T\G_{abcd}\eta,
\ee
\be
\Om_{1\,abc}=-i\eta^T\G_{abc}\eta,\qquad\Om_{2\,abc}=-\eta^T\G_{abc}\G\eta.
\ee
It is easy to verify that these forms satisfy the duality relations
\be
\ast J=L,\qquad\ast L=J,\qquad\ast\Om_1=-\Om_2,\qquad\ast\Om_2=\Om_1.
\ee
Note that $\ast$ squares to one on even forms but to minus one on odd forms.  Using (\ref{eq:FirstFierz}) and manipulations with epsilon tensors, we can show that the four-form is not independent
\be
L_{abcd}=3J_{[ab}J_{cd]}.
\ee
Because of this, we will have no need to refer to $L_{abcd}$ explicitly in the rest of the note.

Using the Fierz identity (\ref{eq:FirstFierz}) we can derive contraction identities,
\be
\label{eq:ContractionsStart}
J^{ac}J_{bc}=\d^a_b,\qquad J^{ad}\Om_{1\,bcd}=-\Om_{2\,\hph{a}bc}^{\hph{2\,}a},\qquad J^{ad}\Om_{2\,bcd}=\Om_{1\,\hph{a}bc}^{\hph{1\,}a},
\ee
\be
\Om_1^{abe}\Om_{1\,cde}=2\d^{[a}_{[c}\d^{b]}_{d]}-2J^{[a}_{\hph{[a}[c}J^{b]}_{\hph{b]}d]}=\Om_2^{abe}\Om_{2\,cde},\qquad\Om_1^{abe}\Om_{2\,cde}=4\d^{[a}_{[c}J^{b]}_{\hph{b]}d]}.
\ee
\be
J^{ab}J_{ab}=6,\quad J^{bc}\Om_{1\,abc}=J^{bc}\Om_{2\,abc}=0,\quad\Om_1^{acd}\Om_{1\,bcd}=\Om_2^{acd}\Om_{2\,bcd}=4\d^a_b,
\ee
\be
\label{eq:ContractionsEnd}
\Om_1^{acd}\Om_{2\,bcd}=4J^a_{\hph{a}b},\qquad\Om_1^{abc}\Om_{1\,abc}=\Om_2^{abc}\Om_{2\,abc}=24,\qquad\Om_1^{abc}\Om_{2\,abc}=0.
\ee
Similarly we can obtain the orthogonality relations and normalization constraints
\be
\label{eq:Constraints}
J_{[ab}\Om_{1\,cde]}=J_{[ab}\Om_{2\,cde]}=0,\qquad\Om_{1\,[abc}\Om_{2\,def]}=3J_{[ab}J_{cd}J_{ef]}.
\ee



In terms of maps between variables, a choice of forms $\{J,\Om_1\}$, satisfying  (\ref{eq:Constraints})
is equivalent to the metric and normalized spinor, $\{g,\eta\}$. Here $\Om_2$ is constructed from $J$ and $\Om_1$, as we discuss in the next subsection.
As a check, we have the counting
\be
\lp\#\ \mathrm{of\ components\ of\ }\{g,\eta\}\rp=21+7=28,
\ee
\begin{multline}
\lp\#\ \mathrm{of\ components\ of\ }\{J,\Om_1,\Om_2\}\rp-\lp\#\ \mathrm{of\ constraints}\rp\\
=\lp 15+20\rp-\lp 6+1\rp=28,
\end{multline}
and
\be
\dim\lp\operatorname{GL}(6)\rp-\dim\lp\SU(3)\rp=36-8=28.
\ee

\subsection{Metric from $\SU(3)$-structure}

Suppose we are given a real two-form $J$ and a real three-form $\widetilde{\Om}_1$ satisfying $J\w\widetilde{\Om}_1=0$.
The tilde is because $\Om_1$ is related to $\widetilde{\Om}_1$ by the rescaling below.
Using the antisymmetric symbol (not tensor) $\hat{\e}$, define a symmetric matrix in terms of the $SU(3)$-structure,
\be
s_{ab}=\hat{\e}^{cdefgh}\widetilde{\Om}_{1\,acd}\widetilde{\Om}_{1\,bef}J_{gh}.
\ee
Let $s=\det(s_{ab})$ and as long as $s\ne 0$ define
\be
\widetilde{g}_{ab}=\hlf s^{-\frac{1}{8}}s_{ab}.
\ee

Now define a scaling factor
\be
\la=\frac{1}{6}\widetilde{g}^{ab}\widetilde{g}^{cd}J_{ac}J_{bd},
\ee
and finally we obtain the three form and metric,
\be
\Om_{1\,abc}=\la\widetilde{\Om}_{1\,abc},\qquad g_{ab}=\la^{1/2}\widetilde{g}_{ab}.
\ee
From (\ref{eq:FirstFierz}) we see that $\Om_{2\,abc}$ is not independent
\be
\Om_{2\,abc}=-J_{ad}g^{de}\Om_{1\,bce}.
\ee
With these expressions we can verify that the various contractions (\ref{eq:ContractionsStart})-(\ref{eq:ContractionsEnd}) are satisfied.

It will be useful to have the relations above to linear order in perturbations around a background solution given by $\{J,\Om_1,\Om_2,g\}$.  Suppose we are given perturbations $\d J_{ab}$ and $\d\Om_{1\,abc}$.  To ensure that the constraints are satisfied, these must obey (here indices are raised and lowered with the uncorrected metric)
\be
\label{eq:6DLinearizedConstraints}
J^{bc}\d\Om_{1\,abc}+\Om_{1\,a}^{\hph{1\,a}bc}\d J_{bc}=0,\qquad 6J^{ab}\d J_{ab}-\Om_1^{abc}\d\Om_{1\,abc}=0.
\ee
The first condition guarantees $J\w\Om_1=0$ and the second $J\w J\w J=\frac{3}{2}\Om_1\w\Om_2$, even once the perturbations are included.  As long as these relations are obeyed by $\d J_{ab}$ and $\d\Om_{1\,abc}$, then we have
\bea
\d g_{ab} &=& -\hlf g_{ab}J^{cd}\d J_{cd}+J_{(a}^{\hph{(a}c}\d J_{b)c}+\hlf\Om_{1\,(a}^{\hph{1\,(a}cd}\d\Om_{1\,b)cd},\\
\label{eq:6DDeltaOm2}
\d\Om_{2\,abc} &=& -J_a^{\hph{a}d}\d\Om_{1\,bcd}-\Om_{1\,bc}^{\hph{1\,bc}d}\d J_{ad}+J_a^{\hph{a}d}\Om_{1\,bc}^{\hph{1\,bc}e}\d g_{de}.
\eea

\subsection{Decomposition of forms}
Similarly as we did for $n=4$, we present the decomposition of forms into $SU(3)$ representations and derive some useful
formulas needed to evaluate the gravitino SUSY transformation.
Under $\SO(6)\rr\SU(3)$, the spaces of differential forms decompose as
\bea
\La^0 &=& \La^0_1,\\
\La^1 &=& \La^1_{3\oplus\bar{3}},\\
\La^2 &=& \La^2_1\oplus\La^2_{3\oplus\bar{3}}\oplus\La^2_8,\\
\La^3 &=& \La^3_{1\oplus 1}\oplus\La^3_{3\oplus\bar{3}}\oplus\La^3_{6\oplus\bar{6}},\\
\La^4 &=& \La^4_1\oplus\La^4_{3\oplus\bar{3}}\oplus\La^4_8,\\
\La^5 &=& \La^5_{3\oplus\bar{3}},\\
\La^6 &=& \La^6_1.
\eea
At any given point the singlet spaces $\La^0_1$, $\La^2_1$, $\La^3_{1\oplus 1}$, $\La^4_1$, and $\La^6_1$ are spanned by $1$, $J$, $\Om_1$ and $\Om_2$, $J\w J$, and $J\w J\w J$, respectively.

We can derive projectors for the two-, three-, and four-forms.  For a two-form $\al_{ab}$, we have
\bea
\pi_1(\al)_{ab} &=& \frac{1}{6}J_{ab}J^{cd}\al_{cd},\\
\pi_{3\oplus\bar{3}}(\al)_{ab} &=& \hlf\al_{ab}-\hlf J_{[a}^{\hph{[a}c}J_{b]}^{\hph{b]}d}\al_{cd},\\
\pi_8(\al)_{ab} &=& \hlf\al_{ab}+\hlf J_{[a}^{\hph{[a}c}J_{b]}^{\hph{b]}d}\al_{cd}-\frac{1}{6}J_{ab}J^{cd}\al_{cd},
\eea
while the remaining projectors are written for completeness in the appendix. The former projector is all we need in the following.

Next we would like to write an arbitrary one-form on a Calabi-Yau threefold, along the lines of (\ref{eq:OneFormComponents}). On a Calabi-Yau three-fold there are no harmonic one-forms, so the Hodge decomposition for one-forms is
\be
\xi=d\la+d^\dagger\beta.
\ee
To see that we can make a convenient choice for $\beta$ note that we have the following equation for one-forms $\al$,
\be
d^\dagger\lp\pi_8-\pi_{3\oplus\bar{3}}-2\pi_1\rp d\al=0,
\ee
which follows by considering the explicit form of the projection operators.  Then we can do a Hodge decomposition of $\beta$,
\be
\beta=d\al+d^\dagger\om+\g,
\ee
where $\g$ is harmonic.  Let
\be
\beta'=\lp 3\pi_1+2\pi_{3\oplus\bar{3}}\rp d\al=\beta-\lp\pi_8-\pi_{3\oplus\bar{3}}-2\pi_1\rp d\al-d^\dagger\om-\g.
\ee
Then clearly we have $\pi_8\beta'=0$ and $d^\dagger\beta'=d^\dagger\beta$. Thus we can always assume that there is no $8$ piece in $\beta$. The decompositions then take the explicit form
\be
\beta_{ab}=xJ_{ab}+\Om_{1\,abc}v^c,
\ee
and
\be
\label{eq:6Ddecomposition}
\xi_a=\nabla_a\la+J_a^{\hph{a}b}\nabla_bx+\Om_{1\,a}^{\hph{1\,a}bc}\nabla_bv_c.
\ee
We can actually go further. Since only $dv$ appears above, we can always shift $v_a$ by something exact to arrange that $v$ is co-closed, $\nabla^av_a=0$.  Similarly, we can show that it is always possible to shift $v$ to arrange that $J\cdot v$ is also co-closed, i.e. that $J^{ab}\nabla_av_b=0$.

\subsection{Gravitino SUSY variation and background geometry}

The most general form of the gravitino SUSY transformation expresses $\eta$ in terms of the basis of spinors
\be
\nabla_a\eta=iA_a\G\eta+iB_{ab}\G^b\eta.
\ee
Here $A_a$ and $B_{ab}$ may involve a derivative expansion in terms of Riemann tensors, non-perturbative effects or/or fluxes. E.g.\ it is known that the $\al'^3$ perturbative contribution is non-vanishing for compactifications on Ricci flat K\"ahler manifolds (see \cite{Grisaru:1986dk}, \cite{Nemeschansky:1986yx}, \cite{Gross:1986iv} and \cite{Candelas:1986tz}).
From the previous equation we can compute the derivatives for the $SU(3)$-structure
\bea
\nabla_aJ_{bc} &=& -2i\eta^T\G_{bc}\G\nabla_a\eta=2B_a^{\hph{a}d}\Om_{2\,bcd},\\
\nabla_a\Om_{1\,bcd} &=& -2i\eta^T\G_{bcd}\nabla_a\eta=-2A_a\Om_{2\,bcd}-2B_a^{\hph{a}e}L_{bcde},\\
\nabla_a\Om_{2\,bcd} &=& -2\eta^T\G_{bcd}\G\nabla_a\eta=2A_a\Om_{1\,bcd}-6B_{a[b}J_{cd]},\\
\eea
Antisymmetrizing, this gives $dJ$, $d\Om_1$ and $d\Om_2$, in terms of $A_a$ and $B_{ab}$.  In general we can expand these forms in their $\SU(3)$ representations. These are not all independent, because of the relations obeyed by the underlying forms $J$, $\Om_1$, and $\Om_2$.  The independent components transform as
\be
\ls 2\times\mathbf{1}\rs\oplus\ls 2\times\lp\mathbf{3}\oplus\bar{\mathbf{3}}\rp\rs\oplus\lp\mathbf{6}\oplus\bar{\mathbf{6}}\rp\oplus\ls 2\times\mathbf{8}\rs,
\ee
which is a subset of the representations in $\La^3\oplus\La^4\oplus\La^4$.  These are also known as the $\SU(3)$-structure torsion classes $W_i$, and they are in one-to-one correspondence with the forty-two components of $A_a$ and $B_{ab}$.  Note that $\al$, $\beta$, and $\g$ are closed forms by construction. This follows by using the integrability condition for the SUSY variation.

Given $A_a$ and $B_{ab}$ and an uncorrected solution for $J$, $\Om_1$ and $\Om_2$ satisfying the constraints
(\ref{eq:Constraints}), our goal is to find $J'=J+\d J$ and $\Om_1'=\Om_1+\d\Om_1$ such that
\be
dJ'=\al,\qquad d\Om_1'=\beta,\qquad d\Om_2'=\g,\non
\ee
\be
J'\w\Om_1'=0,\qquad J'\w J'\w J'=\frac{3}{2}\Om_1'\w\Om_2',
\ee
Here $\Om_2'$ is derived from $J'$ and $\Om_1'$ and $\al$, $\beta$ and $\g$ are constructed from $A_a$ and $B_{ab}$, as we show for the $B_{ab}=0$ case next.

Take $B_{ab}=0$ and interpret $A_a$ as a one-form.  In this case the corrected two form $J'$ is closed, while the $d\Om'$ components receive a correction,
\be
\al=0,\qquad\beta=-2 A\w\Om_2,\qquad\g=2 A\w\Om_1.
\ee
As a result the corrected manifold remains K\"ahler. To work out the explicit expression for the one form $A$ notice that
the integrability conditions say
\be
dA\w\Om_1=dA\w\Om_2=0.
\ee

%

Taking this into account and using the decomposition (\ref{eq:6Ddecomposition}) for $A$, we derive the expression
\be
A_a=\nabla_a\la_A+J_a^{\hph{a}b}\nabla_bx_A.
\ee
Just as in four dimensions, the integrability conditions require the term with $v$ in the expansion of $A_a$ to vanish.

Similarly as in 4D this points to an ansatz
\be
J'=J+da,\qquad\Om_1'=\Om_1-2\la_0\Om_2+2x_0\Om_1+db.
\ee
In components,
\be
\d J_{ab}=2\nabla_{[a}a_{b]},\qquad\d\Om_{1\,abc}=-2\la_A\Om_{2\,abc}+2x_A\Om_{1\,abc}+3\nabla_{[a}b_{bc]}.
\ee
The goal is now to pick $a_a$ and $b_{ab}$ such that the linearized constraints (\ref{eq:6DLinearizedConstraints}) are obeyed and
\be
\label{eq:6DdOm2Linearized}
4\nabla_{[a}\d\Om_{2\,bcd]}=-8\Om_{1\,[abc}\nabla_{d]}\la_A-8\Om_{2\,[abc}\nabla_{d]}x_A,
\ee
where $\d\Om_{2\,abc}$ is defined by (\ref{eq:6DDeltaOm2}).

First we will exhibit a particular solution to this system, by taking $b_{ab}=0$ and $a_a=J_a^{\hph{a}b}\nabla_b\rho$, for some function $\rho$.  Then it is easy to check that the first constraint of (\ref{eq:6DLinearizedConstraints}) is satisfied, while the second becomes a Poisson equation
\be
\label{eq:6DParticularPoisson}
\nabla^2\rho+4x_A=0.
\ee
As is familiar from the four-dimensional case, we can always solve this condition, with $x_A$, possibly shifted by a constant, acting as a source term.

Now we compute
\bea
\d g_{ab} &=& -\nabla_{(a}\nabla_{b)}\rho-J_{(a}^{\hph{(a}c}J_{b)}^{\hph{b)}d}\nabla_c\nabla_d\rho,\\
\d\Om_{2\,abc} &=& 2\la_A\Om_{1\,abc}+2x_A\Om_{2\,abc},
\eea
where in both cases we have used our constraint equation (\ref{eq:6DParticularPoisson}).  It is now simple to check that (\ref{eq:6DdOm2Linearized}) is also satisfied, so we have a solution to the full system of equations.

In fact, we can exhibit a whole six-dimensional space of solutions.  Let $v^a$ be an arbitrary vector on the manifold, and take
\be
a_a=J_a^{\hph{a}b}\nabla_b\rho-J_{ab}v^b,\qquad b_{ab}=\Om_{1\,abc}v^c.
\ee
Then we can check that $v^a$ drops out of the constraints (\ref{eq:6DLinearizedConstraints}), so we have only the Poisson equation (\ref{eq:6DParticularPoisson}) for $\rho$ again.  Then we also have
\bea
\d g_{ab} &=& -\nabla_{(a}\nabla_{b)}\rho-J_{(a}^{\hph{(a}c}J_{b)}^{\hph{b)}d}\nabla_c\nabla_d\rho+\nabla_av_b+\nabla_bv_a,\\
\d\Om_{2\,abc} &=& 2\la_A\Om_{1\,abc}+2x_A\Om_{2\,abc}-\Om_{2\,[ab}^{\hph{2\,[ab}d}\nabla_{c]}v_d.
\eea
Since the term added to $\d\Om_2$ is exact, it of course drops out of the remaining equation (\ref{eq:6DdOm2Linearized}), and so we still have a solution to the full system of equations.  Moreover, from the form of the addition to $\d g_{ab}$, we see that $v^a$ acts simply as an infinitesimal diffeomorphism.  We expect this to be the full space of solutions.

\section*{Acknowledgement} This work
was supported by the grants PHY-1214344 and NSF Focused Research Grant DMS-1159404 and by
the George P. and Cynthia W. Mitchell Institute for Fundamental Physics and Astronomy.
We thank J.~Caldeira, A.~Royston, and S.~Sethi for useful discussions.

\appendix

\section{Useful formulae}

For completeness we have collected some additional formulas for the $n=6$ case in this appendix.
\begin{enumerate}

\item To make contact with the approach using complex spinors, note that we can define a complex chiral spinor by
\be
\label{eq:ZetaDef}
\zeta=\frac{1}{\sqrt{2}}\lp 1+\G\rp\eta.
\ee
This satisfies
\be
\G\zeta=\zeta,\qquad\G\overline{\zeta}=-\overline{\zeta},\qquad\zeta^\dagger\zeta=1.
\ee
Moreover, using the contraction identities we can show that
\be
\lp i\G_a-J_a^{\hph{a}b}\G_b\rp\zeta=0,\qquad\lp i\G_a+J_a^{\hph{a}b}\G_b\rp\overline{\zeta}=0.
\ee
If we picked holomorphic coordinates by splitting into $+i$ and $-i$ eigenspaces of the matrix $J_a^{\hph{a}b}$ (which squares to minus one), then this states that
\be
\G_{\bar{\imath}}\zeta=0,\qquad\G_i\overline{\zeta}=0.
\ee

Conversely, starting with a nowhere-vanishing complex chiral spinor $\zeta$, which we can normalize so that $\zeta^\dagger\zeta=1$, we can define a real spinor
\be
\eta=\sqrt{2}\operatorname{Re}(\zeta),
\ee
which implies (\ref{eq:ZetaDef}). In particular $\eta$ is also nowhere-vanishing.

\item The projectors for the three-form $\beta_{abc}$ are,
\bea
\pi_{1\oplus 1}(\beta)_{abc} &=& \frac{1}{24}\lp\Om_{1\,abc}\Om_1^{def}\beta_{def}+\Om_{2\,abc}\Om_2^{def}\beta_{def}\rp,\\
\pi_{3\oplus\bar{3}}(\beta)_{abc} &=& \frac{3}{4}J_{[ab}J^{de}\beta_{c]de},\\
\label{eq:Pi66}
\pi_{6\oplus\bar{6}}(\beta)_{abc} &=& \beta_{abc}-\frac{3}{4}J_{[ab}J^{de}\beta_{c]de}\non\\
&& \qquad -\frac{1}{24}\lp\Om_{1\,abc}\Om_1^{def}\beta_{def}+\Om_{2\,abc}\Om_2^{def}\beta_{def}\rp.
\eea

For a four-form $\g_{abcd}$ the projectors take the form,
\bea
\pi_1(\g)_{abcd} &=& -\frac{1}{3}\g_{abcd}+J_{[ab}J^{ef}\g_{cd]ef},\\
\pi_{3\oplus\bar{3}}(\g)_{abcd} &=& \g_{abcd}-3J_{[a}^{\hph{[a}e}J_b^{\hph{b}f}\g_{cd]ef},\\
\pi_8(\g)_{abcd} &=& \frac{1}{3}\g_{abcd}-J_{[ab}J^{ef}\g_{cd]ef}+3J_{[a}^{\hph{[a}e}J_b^{\hph{b}f}\g_{cd]ef}.
\eea

\end{enumerate}

\newpage

\providecommand{\href}[2]{#2}\begingroup\raggedright
\endgroup

\end{document}